%% file: 31645_corr_v2.tex
\documentclass[printer]{aa}
\usepackage[utf8]{inputenc}%
\usepackage[english]{babel}%
\usepackage[dvips]{graphicx}%
\usepackage{amssymb}
\usepackage{hhline}

\begin{document}

\title{Cosmic flow around local massive galaxies}
\titlerunning{Cosmic flow around local massive galaxies}

\author{Olga G. Kashibadze\inst{1}\fnmsep\thanks{f.k.a. Nasonova.}\and
        Igor D. Karachentsev\inst{1}}
\authorrunning{O.\,G.\,Kashibadze \& I.\,D.\,Karachentsev}

\institute{Special Astrophysical Observatory of the Russian Academy of Sciences\\
                                                              \\
          Nizhnij Arkhyz, Karachay-Cherkessia, 369167, Russia}

\date{\today}

\abstract
{}
{We use accurate data on distances and radial velocities of galaxies around the
Local Group, as well as around 14 other massive nearby groups, to estimate their
radius of the zero-velocity surface, $R_0$, which separates any group against
the global cosmic expansion.}
{Our $R_0$ estimate was based on fitting the data to the velocity field expected
from the spherical infall model, including effects of the cosmological constant.
The reported uncertainties were derived by a Monte Carlo simulation.}
{Testing various assumptions about a location of the group barycentre, we
found the optimal estimates of the radius to be $0.91\pm0.05$~Mpc for the Local
Group, and $0.93\pm0.02$~Mpc for a synthetic group stacked from 14 other groups
in the Local Volume. Under the standard Planck model parameters, these
quantities correspond to the total mass of the group $\sim (1.6\pm0.2) 10^{12}
M_{\sun}$. Thus, we are faced with the paradoxical result that the total mass
estimate on the scale of $R_0 \approx (3- 4) R_{vir}$ is only $~60$\% of the
virial mass estimate. Anyway, we conclude that wide outskirts of the nearby
groups do not contain a large amount of hidden mass outside their virial radius.}
{}


\maketitle

\section{Introduction}
Any overdense region in the Universe is driven by the competition between its
self-gravity and the cosmic expansion, and therefore can be characterized by an
idealized zero-velocity surface that separates these zones. De Vaucouleurs
(1958, 1964, 1972) presupposed systematic deviations from linearity in the
velocity-distance relation and interpreted these deviations as a local
phenomenon caused by the Virgo complex. The expected effect has only
subsequently been supported by observations. Peebles (1976) found the
virgocentric infall signal using the field galaxy data available at that time
(Sandage \& Tammann 1975).

Lynden-Bell (1981) and Sandage (1986) focussed on the Local Group of galaxies.
They showed that, in the simplest case of the spherically symmetric system in
the empty Universe with $\Lambda=0$, the radius of the zero-velocity surface
$R_0$ and the total mass of the group $M^0_T$ are related as

\begin{equation} 
M^0_T= (\pi^2/8G)\times R^3_0\times T^{-2}_0,
\end{equation}
where $G$ is the gravitational constant and $T_0$ is the age of the Universe
(Lynden-Bell 1981, Sandage 1986). In the standard cosmological $\Lambda$CDM
model, where $\Omega_m$ is the mean cosmic density of matter and $H_0$ the
Hubble parameter, the relation between $R_0$ and $M_T$ becomes

\begin{equation}
M_T=(\pi^2/8G)\times R^3_0\times H^2_0/f^2(\Omega_m),
\end{equation}
where the dimensionless parameter

\begin{equation}
f(\Omega_m)=(1-\Omega_m)^{-1}-\frac{\Omega_m}{2} (1-\Omega_m)^{-\frac{3}{2}} \cosh^{-1}(\frac{2}{\Omega_m}-1) 
\end{equation}
changes in the range from 1 to 2/3 while varying $\Omega_m$ from 0 to 1. Taking
the Planck model parameters $\Omega_m =0.315, \Omega_{\lambda} = 0.685$ and
$H_0$ = 67.3 km\,s$^{-1}$\,Mpc$^{-1}$ (Planck Collaboration 2014), we obtain the
relation

\begin{equation}
(M_T/M_{\odot})_{0.315} = 1.95\times 10^{12} (R_0/Mpc)^3,
\end{equation}
which is by 1.50 times more than the classical estimate from equation (1).

This method was sucessfully applied to determine masses of the Local Group
(Ekholm et al. 2001, Karachentsev et al. 2002, Teerikorpi et al. 2005,
Karachentsev et al. 2009), M81 group (Karachentsev \& Kashibadze 2006), CenA
group (Karachentsev et al. 2006), as well as the Virgo cluster (Tully \& Shaya
1984, Karachentsev \& Nasonova 2010, Karachentsev et al. 2014) and the Fornax
cluster (Nasonova et al. 2011).

It is important to stress that the $R_0$ method estimates the total mass of a
group independently of mass estimates based on virial motions. Notably, the
corresponding total mass $M_T$ is confined on the linear scale of $R_0$, which
is three to four times as large as the virial radius of a group or cluster,
$R_{vir}$.

\begin{figure*} 
\centering
\includegraphics[width=17cm]{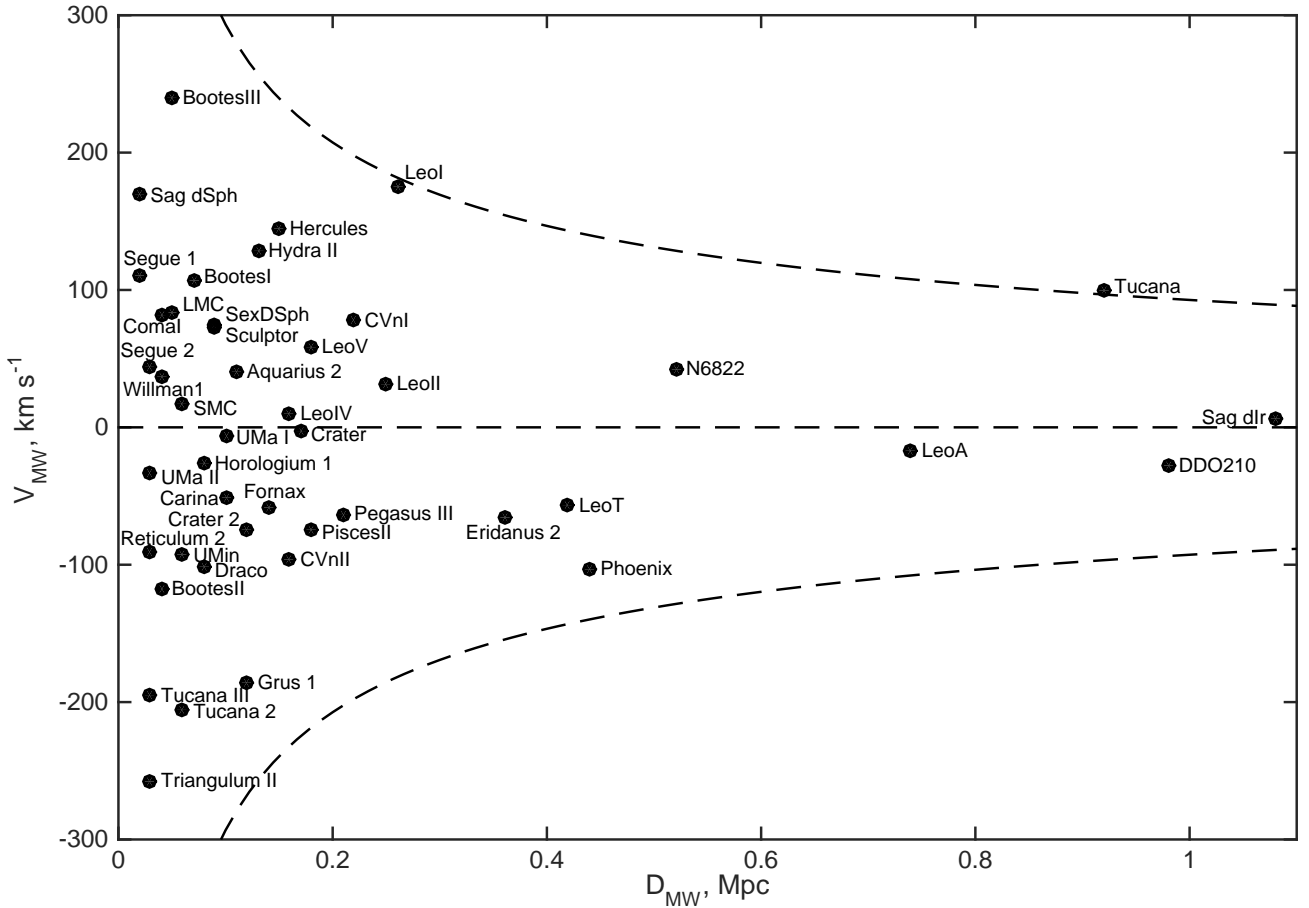}
\caption{Distribution of 45 Milky Way satellites by their spatial distances and
radial velocities relative to the Milky Way. Dashed lines correspond to the
parabolic velocity for a point mass of $1\times10^{12}M_{\odot}$.}
\label{fig1}
\end{figure*}

The implementation of the $R_0$ method became possible with wholesale
measurements of distances to nearby galaxies from luminosities of the red giant
branch stars (TRGB) with accuracy of $\sim5$\% attainable by the Hubble Space
Telescope. In the Local Volume, limited to 11 Mpc, there are about a thousand
known galaxies; most of these galaxies have measured radial velocities with a
typical accuracy less than 5 km\,s$^{-1}$. About one-third of the Local Volume
population already has accurate TRGB distance estimates. The compilation of
observational data on these objects is presented in the Updated Nearby Galaxy
Catalog (Karachentsev et al. 2013) and its latest electronic version:
http://sao.ru/lv/lvgdb/ (Kaisina et al. 2012). For a typical galaxy of the Local
Volume with a distance of $\sim6$ Mpc, the TRGB distance error of $\sim300$ kpc
is comparable with a virial radius of the group, thus its location can be
confidently fixed relative to the group centroid and zero velocity surface.
Other methods of secondary importance are the Tully \& Fisher (1977) relation
distances or the brightest stars distances with an accuracy of $\sim(20-30)$\%
do not provide an opportunity to determine $R_0$ value even for the nearest
groups.

Below we use the most complete data on distances and radial velocities of the
Local Volume galaxies to estimate the zero-velocity radius around the local 
massive galaxies.

\section{Galaxy motions around the Milky Way and M31}

\begin{figure*} 
\centering
\includegraphics[width=17cm]{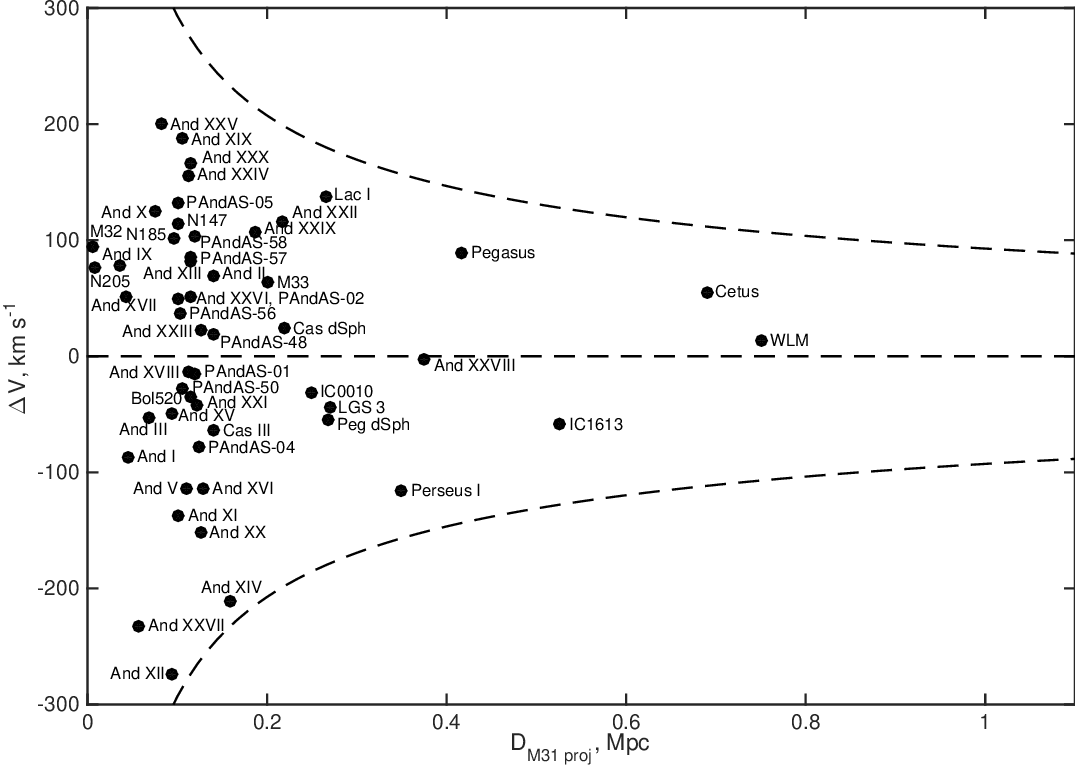}
\caption{Distribution of 52 test particles by their differential radial
velocities and projected distances from M31. Dashed lines correspond to the
parabolic velocity for a point mass of $1\times10^{12}M_{\odot}$.}
\label{fig2}
\end{figure*}

The recent surveys of large sky areas (Abazajian et al. 2009, Tonry et al. 2012,
Koposov et al. 2015) led to the discovery of new Milky Way (MW) dwarf
satellites with low luminosities and extremely low surface brightnesses. The
recent overview by McConnachie (2012) reports 29 MW satellites with measured
radial velocities and accurate distances. In recent years, this list has
been expanded up to 45 objects. The corresponding data are presented in Table~1.
The Table columns contain (1) galaxy name; (2) equatorial coordinates
J2000.0; and (3) tidal index,

\begin{equation} 
TI = \max[\log(M^*_n/D^3_n)]-10.96, \,\,\,\, n=1, 2, \ldots N,
\end{equation}
distinguishing the most significant galaxy (main disturber = MD) among $N$
neighbouring galaxies, whose tidal force dominates the remaining galaxies with
masses $M^*_n$ and spatial separations $D_n$. The constant, --10.96, is chosen
in such a way that $TI=0$ corresponds to a significant neighbour located on the
zero velocity surface with $TI<0$ galaxies ranked as isolated. Finally, Col. (4)
lists the main disturber name, Col. (5) distance to a galaxy in Mpc, and Cols.
(6, 7) radial velocity of a galaxy (in km\,s$^{-1}$) relative to the Sun and
relative to the Milky Way centre with apex parameters adopted in NASA
Extragalactic Database (NED). References to the used values of distances and
velocities of galaxies are presented in the Local Volume Galaxies Database
(http://sao.ru/lv/lvgdb/).

The distribution of 45 satellites of the MW by their Galactocentric distances
and radial velocities is shown in Figure~1. The dashed lines correspond to the
parabolic velocity for a point mass of $1\times10^{12}M_{\odot}$. The velocity
distribution of satellites looks symmetrical relative to the MW centre, although
two satellites with near-parabolic velocities~--- Tucana and LeoI~--- are close
to the upper escape limit. Three MW satellites, Sag~dIr, DDO\,210, and Tucana
with distances $D\sim 1$~Mpc and negative $\Theta_1$, belong to field galaxies.
However, the MW is dynamically the most significant neighbour for each of these.
   
Specialized searches for faint satellites in the outskirts of the spiral galaxy
M31 in the Andromeda constellation (Ibata et al. 2007, Martin et al. 2009, Ibata
et al. 2014) has proved to be notably productive. While the sample by
McConnachie (2012) included 23 satellites, now their number is roughly doubled
amounting up to 44. The data on these satellites are presented in Table~2, where
the first six columns have the same meaning as in Table~1. The seventh column of
Table~2 contains spatial distances of satellites (in Mpc) relative to M31, while
eighth and ninth list the projected separation of satellites in the sky (in Mpc)
and their differential radial velocities relative to M31 (in km\,s$^{-1}$).
Aside from dwarf galaxies, we tabulate also the data on eight distant globular
clusters from PAndAS survey (Huxor et al. 2014) with measured radial velocities.
Their spatial distances still remain unknown, and we set them equal to 0.78 Mpc.

The distribution of $44+8$ test particles by their differential radial
velocities and projected separations relative to M31 is presented in Figure~2.
The dashed lines also mean the parabolic velocity for a point mass of
$1\times10^{12}M_{\odot}$. Similar to the MW case, the distribution of M31
satellites by their relative velocities seems to be very symmetrical; two
satellites~--- And~XIV and And~XII~--- have near-parabolic velocities that are
close to the lower escape limit. 

\section{Orbital masses of the Milky Way and M~31}

For a massive galaxy surrounded by small satellites, the orbital mass estimate
is expressed as

\begin{equation}
M_{orb}=(32/3\pi)\times(1-2e^{2}/3)^{-1}\times G^{-1}\times\langle\Delta V^2\times R_p\rangle,
\end{equation}
where $\langle\Delta V^2 \times R_p\rangle$ is the mean product of squared
radial velocity difference of a satellite with its projected distance from the
main galaxy and $e$ is the orbit eccentricity (Karachentsev \& Kudrya 2014).
This relation is obtained under the assumption of uniformly random orientation
of satellite orbits relative to the line of sight. With the typical eccentricity
value of $\langle e^2\rangle = 1/2$ (Barber et al. 2014) the relation (6)
becomes 

\begin{equation}
M_{orb}=(16/\pi)\times G^{-1}\times \langle\Delta V^2 \times R_p\rangle.
\end{equation}

Applying eq. (7) to the assembly of the MW and M31 satellites, we get values for
orbital masses $M_{orb}$(MW)$ = 1.51\times10^{12}M_{\odot}$ and $M_{orb}$
(M31)$ = 1.69\times10^{12}M_{\odot}$. Since in the case of Milky Way satellites
we observe their 3D distances, then project distances, the orbital mass
estimation should be reduced by a factor of $(\pi/4)$ yielding $M_{orb}$(MW)$ =
1.18\times10^{12}M_{\odot}$. Hence, the ratio of mass estimates for these two 
galaxies reaches

\begin{equation}
M_{orb}(MW)/M_{orb}(M31)\simeq0.70.
\end{equation}

This value is quite close to the ratio $M_{orb}(MW)/M_{orb}$(M31)=0.82 obtained
by Karachentsev \& Kashibadze (2006) from a minimum value for scatter of
peculiar velocities with respect to the Hubble regression line, while varying
the centroid position between the MW and M31.

A comparison of the derived total masses of the MW and M31, their combined mass,
and the mass ratio with other mass estimates in the recent literature is
presented in Table 3. These estimates were based on kinematics of satellites and
globular clusters assuming that the MW and M31 haloes follow the standard NFW
profile or fit the kinematics of high-velocity stars and blue horizontal branch
stars.  Our present measurements are in good agreement with the median values
given in the last line of Table 3. An essential part of the mismatch between the
different estimates in Table 3 may arise from the observed orbital anisotropy of
the MW and M31 satellites (Ibata et al. 2013, Pawlowski et al. 2014) and from
the uncertain dynamical status of two Milky Way satellites, Leo~I and Tucana,
and  the two M31 satellites, And~XIV and And~XII. Excluding these objects
reduces the mass estimates by 14--15\% in both cases.

\section{Hubble flow around the Local Group}

The proximate velocity field around the Local Group was considered in most
detail by Karachentsev \& Kashibadze (2006) and Karachentsev et al. (2009). For
a sample of 30 galaxies with TRGB distances from 0.7 to 3.0 Mpc with respect to
the Local Group centre, it was shown that the Hubble flow is characterized by
the local Hubble parameter $H_{loc}=(78\pm2)$ km\,s$^{-1}$ Mpc$^{-1}$, the
radial velocity dispersion $\sigma_v\simeq25$ km\,s$^{-1}$, and the radius of
zero-velocity surface $R_0=(0.96\pm0.03)$ Mpc. The minimal value of $\sigma_v$
corresponded to the barycentre position of $D_c=(0.55\pm0.05) D_{M31}$ = 0.43
Mpc, determining the mass ratio of $M_{MW}/M_{M31}\simeq0.8$ stated above.

\begin{figure} 
\resizebox{\hsize}{!}{\includegraphics{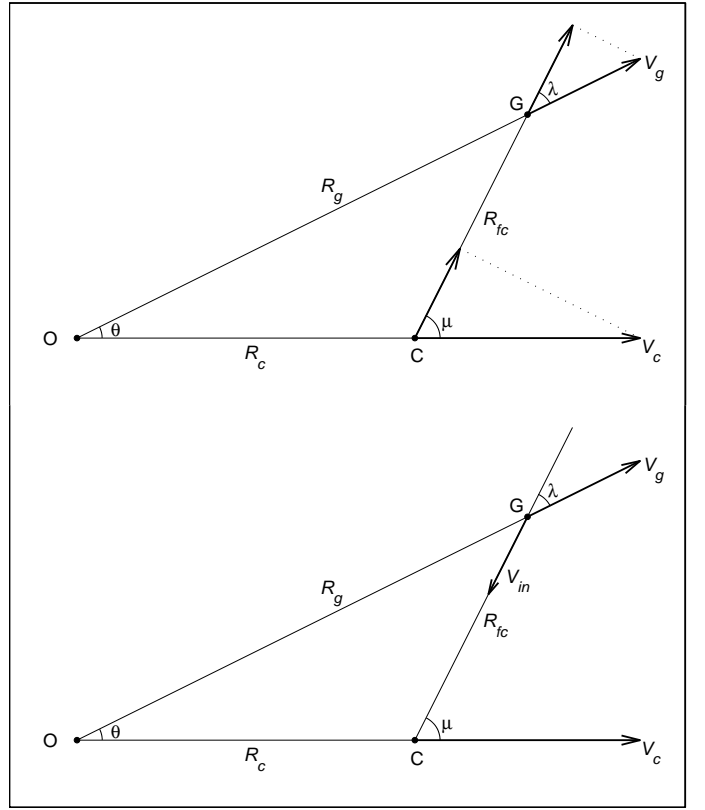}}
\caption{Models of major and minor attractor. O represents an observer, C
represents the centre of a galaxy group, and G represents a test particle (a
galaxy).}
\label{fig3}
\end{figure}

\begin{figure*}
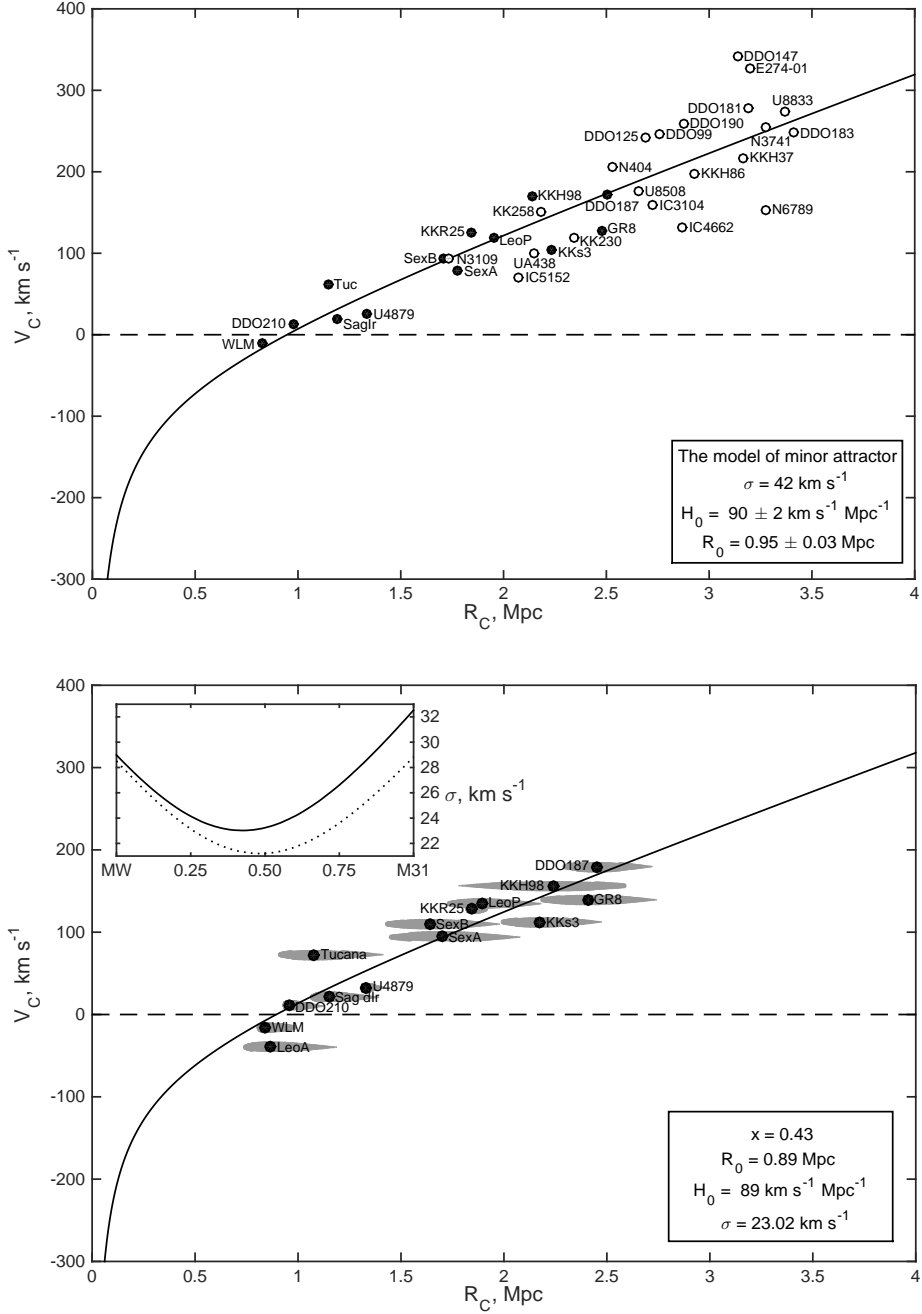

\vspace{0.2cm}%
\includegraphics[width=12cm]{synthetic_regression_lg_binary_minor.eps}\\%
\vspace{0.2cm}\\%
\sidecaption%
\includegraphics[width=12cm]{LGcentroid_lg.eps}%
\caption{Distribution of isolated galaxies by distances and velocities relative
to the Local group barycentre assuming the minor attractor model.
{\itshape{}Upper panel:} The barycentre position locates at $x = D_c/D_{M31} =
0.55$ towards the M 31. The solid circles indicate galaxies that have the MW or
M31 as the main disturber. {\itshape{}Lower panel:} The Hubble flow around the
LG at the barycentre position of $x = D_c/D_{M31} = 0.43$. The grey wedges trace
the companion positions under different $x$, where their thin end  corresponds
to the barycentre coinciding with M~31.Solid and dotted lines in the insert
indicate the velocity dispersion as a function of $x$ with and without the
Leo~A, respectively.\vspace{0.8cm}}
\label{fig4}
\end{figure*}

In that approach, we considered the so-called minor attractor model, illustrated
by the upper panel of Figure~3. Here, a galaxy group with centre, $C,$ is
separated by a distance, $D_c$, from the observer, $O$, and moves away along the
line of site with the velocity, $V_c$. In the outskirts of the group there is a
galaxy, $G$, with distance, $D_g$, and radial velocity, $V_g$. If the angle
between $C$ and $G$ is $\theta$, then their mutual separation is expressed as

\begin{equation}
R^2=D_g^2+D_c^2 -2D_g\times D_c\times\cos \theta,
\end{equation}
and the projected differential velocity is given by

\begin{equation}
V_{gc}=V_g\times\cos\lambda-V_c\times \cos\mu,
\end{equation}
where $\mu=\lambda+\theta$, and

\begin{equation}
\tan\lambda = D_c \times \sin\theta/(D_g-D_c \times \cos\theta).
\end{equation}
In this scheme we assumed peculiar velocities of galaxies in the vicinity
of a group to be small compared with velocities of the regular Hubble flow.

Yet, there is another possibility, which is the major attractor case (see lower
panel of Figure~3); this case is characterized by predominating infall towards
the centre of a group or a cluster. If $V_i$ is the infall velocity than

\begin{equation}
V_g=V_c\times\cos\theta-V_i\times\cos\lambda,
\end{equation}
and the velocity of a galaxy relative to the group centre is expressed as

\begin{equation}
V_i=[V_c\times\cos\theta-V_g]/\cos\lambda.
\end{equation}

Evidently the difference between these two models would be insignificant if the
galaxy lays almost strictly behind ($\lambda\simeq0$) or in front 
($\lambda\simeq180^{\circ}$) of the group centre.
 
The last few years astronomers have detected some new dwarf galaxies in the
vicinity of the Local Group (KKs3, LeoP, and KK258) and measured their accurate
TRGB distances and radial velocities. For some galaxies (KKR25 and Tucana), old
inexact values of radial velocities were corrected and distances were refined.
This circumstance has motivated us to redefine parameters of the local Hubble
flow.
 
To reduce the role of virial motions, we excluded galaxies with $TI>0$ from
consideration; thus, the MW and M31 satellites with distances $D_{MW} <0.8$ Mpc
were consequently excluded. The data on the rest field galaxies with $D_{MW}
<3.5$ Mpc are presented in Table~4. The columns of Table 4 contain (1) galaxy
name; (2) distance (in Mpc) from the Milky Way; (3) heliocentric radial
velocity( in km\,s$^{-1}$); (4) distance from the Local Group barycentre located
at $D_c= 0.43$ Mpc; (5, 6) velocity (in km\,s$^{-1}$) relative to the barycentre
in the case of minor and major attractor, respectively; (7) tidal index; (8) the
main disturber name; and (9) $\lambda$ in degrees (see Figure~3).

The distribution of 35 isolated galaxies by distances and velocities relative to
the Local Group barycentre for the case of minor attractor is presented in the
upper panel of Figure~4. As shown in the Table~4 data, only 14 galaxies of 35
have the MW or M31 as the main disturber; they are denoted by solid circles.
With reference to these objects, the zone affected gravitationally by the Local
Group reaches $R_c\simeq2.5$ Mpc, while more distant field galaxies are
influenced by other massive neighbours of the Local Group, such as M81,
NGC\,253, and NGC\,5128.

According to Peirani \& de Freitas Pacheco (2008), Falco et al. (2014), and
Penarrubia \& Fattahi (2017), the radial velocity profile around the spherically 
symmetrical group or cluster can be expressed as

\begin{equation}
V(R)=H_0\times R-H_0\times R_0\times (R_0/R)^{1/2},
\end{equation}
where $R_0$ is the radius of the zero-velocity surface to be found. The solid
line in Figure~4 corresponds to equation (14) with parameters defined from the
least squares method, $R_0=0.95\pm0.03$ Mpc, $H_0=90\pm2$ km\,s$^{-1}$
Mpc$^{-1}$, and $\sigma_v=42$ km\,s$^{-1}$. The errors of $R_0$ and $H_0$
parameters were estimated using the Monte Carlo method, assuming that
distance errors for galaxies are distributed normally with a typical value of
$\sim5$\%. The peculiar velocity dispersion in the upper panel of Figure~4 is
contributed mostly by distant galaxies, which are disturbed by the neighbouring
groups. Considering the only 14 galaxies in the zone affected gravitationally by
the Local Group, we obtain the following parameters for the surrounding Hubble
flow: $R_0=0.85\pm0.03$ Mpc, $H_0=79\pm3$ km\,s$^{-1}$ Mpc$^{-1}$, and
$\sigma_v=23$ km\,s$^{-1}$. In the major attractor model,
these parameters vary slightly, since $\lambda$ values for these 14 galaxies are
small (see the last column in Table~4).

Three parameters, i.e. $R_0$, $H_0$ and $\sigma_v$, characterizing the local
cosmic expansion, moderately depend on the position of the LG barycentre. Above,
we used the barycentre location at the distance of $D_c= 0.55 D_{M31}$ = 0.43
Mpc, corresponding to the mass ratio of $M_{M31}/M_{MW}$= 1.2. This ratio
matches well with the medians in Table~3. However, Penarrubia et al.(2014) found
that the minimal scatter of nearby galaxies within 3 Mpc around the LG is
achieved with $M_{M31}/M_{MW}$ = 0.75. The authors have concluded that their
analysis rules out models in which M31 is more massive than our Galaxy with
about 95\% confidence. To check this statement, we calculated $\sigma_v$ for 14
nearest isolated galaxies as a function of the position of the LG barycentre $ x
= D_c/D_{M31}$ on the line connecting the MW with M31. The data on $\sigma_v$
and $R_0$ are presented in Table~5. The lower panel of Figure~4 shows the local
Hubble diagram for 14 galaxies at various M31-to-MW mass ratios. Each galaxy is
drawn by grey wedge with caliber inversely related to the dispersion $\sigma_v$
at given barycentre position; thus its thinner end indicates the barycentre
position at M31. The insert in the figure shows the velocity scatter of galaxies
respect to the best-fitting regression line. The solid and dotted lines in the
insert represent the behaviour of $\sigma_v$ for a case of included or excluded
Leo~A, respectively. This dwarf galaxy is a marginally isolated object with the
tidal index TI= +0.03. The derived minimums of these two lines fix the M31-to-MW
mass ratio near 0.7 and 1.0, respectively, not allowing a firm assessment of
which galaxy mass is dominated. Over the range of $M_{M31}/M_{MW}= [1/3 - 3]$
the value of the zero velocity radius is changing within $R_0 =0.86 - 0.96$ Mpc.
Thus, the observed coldness of the local Hubble flow leads us to measure the
radius of the sphere separating the Local Group from the global cosmic expansion
with $\sim5\%$ error. According to (4), the radius $R_0=0.91\pm0.05$ Mpc yields
the total mass estimate  for the Local Group $M_T=(1.5\pm0.2)\times
10^{12}M_{\odot}$ with an unprecedented accuracy, although this quantity lies
below all values of M(MW+M31) in Table 3. The mismatch becomes slightly less
dramatic when the Planck model parameters in (4) replace the WMAP parameters as
follows:\ $\Omega_m =0.24, \Omega_{\lambda} = 0.76$ and $H_0$ = 73
km\,s$^{-1}$\,Mpc$^{-1}$ (Spergel et al. 2007)); this increases the coefficient
in (4) from 1.95 to 2.12.
 
As noted by Chernin et al. (2004), the actual deviation of the binary 
shape of the Local Group from the spherical symmetry produces a minor bias in 
the $R_0$ and mass estimate. According to N-body simulations by Penarrubia 
et al. (2014), neglecting the quadrupole potential overestimates the Local Group 
mass up to $\sim30\%$. 
 
\section{Other massive galaxies in the Local volume}

Considering the Hubble flow around other giant galaxies of the Local Volume, we
selected 15 galaxies with stellar masses $M^*>3\times 10^{10}M_{\odot}$ and
accurate distances. Their overview is presented in Table~6 with objects ranging
by their distances from the observer. For each of these 15 galaxies, surrounded
by a suite of satellites, the second most massive member of its group is also
indicated. In some cases, i.e. M31 and the Milky Way, NGC\,5128 (CenA) and
NGC\,5236, Maffei~2, and IC\,342, the second galaxy is comparable in mass with
the first galaxy and acts itself as the centre of a dynamically separated
subgroup.

The columns of Table~6 contains (1) galaxy name; (2, 3) its supergalactic
coordinates; (4) the galaxy distance from the MW; (5) its radial velocity
relative to the Local Group centroid; (6) logarithmic stellar mass; (7)
logarithmic orbital mass according to Karachentsev \& Kudrya (2014); (8) number
of satellites of the main galaxy with measured radial velocities and accurate
distances.

Aside from the galaxies presented in Table~6, the Local Volume contains another
two massive galaxies~--- NGC\,2903 ($\log M^*=10.82)$ and NGC~6946 ($\log
M^*=10.76)$. But their distances measured from the luminosity of brightest stars
are not yet sufficiently  accurate. In total, the 15 giant galaxies have about
500 satellites in their suites, but, as shown in the last column of Table~6,
only 102 satellites outside the Local Group have accurate estimates of distances
and velocities. Among the second most massive members of 15 groups, three
galaxies~--- NGC\,4242, NGC\,4597 and NGC\,6684~--- have Tully-Fisher distances
with accuracy of $\sim20$\% (denoted with column signs). In 11 of 15 groups, the
main galaxy exceeds its satellites twice or more in mass, allowing us to
estimate its halo mass from the orbital motions. This approach is not worthwhile
in the case of the rich group Leo~I, where NGC\,3379, NGC\,3368, and several
other bright members have compatible luminosities.

Despite the great efforts to measure highly accurate TRGB distances of nearby
galaxies from Hubble Space Telescope data, many neighbouring groups stay still
poorly explored. For example, in the outskirts of giant galaxies NGC\,4594
(Sombrero), NGC\,5055, and  NGC\,3115, no satellites have reliable
distance estimates.
 
\section{Cosmic flow around the synthetic (stacked) nearby group}

\begin{figure*}
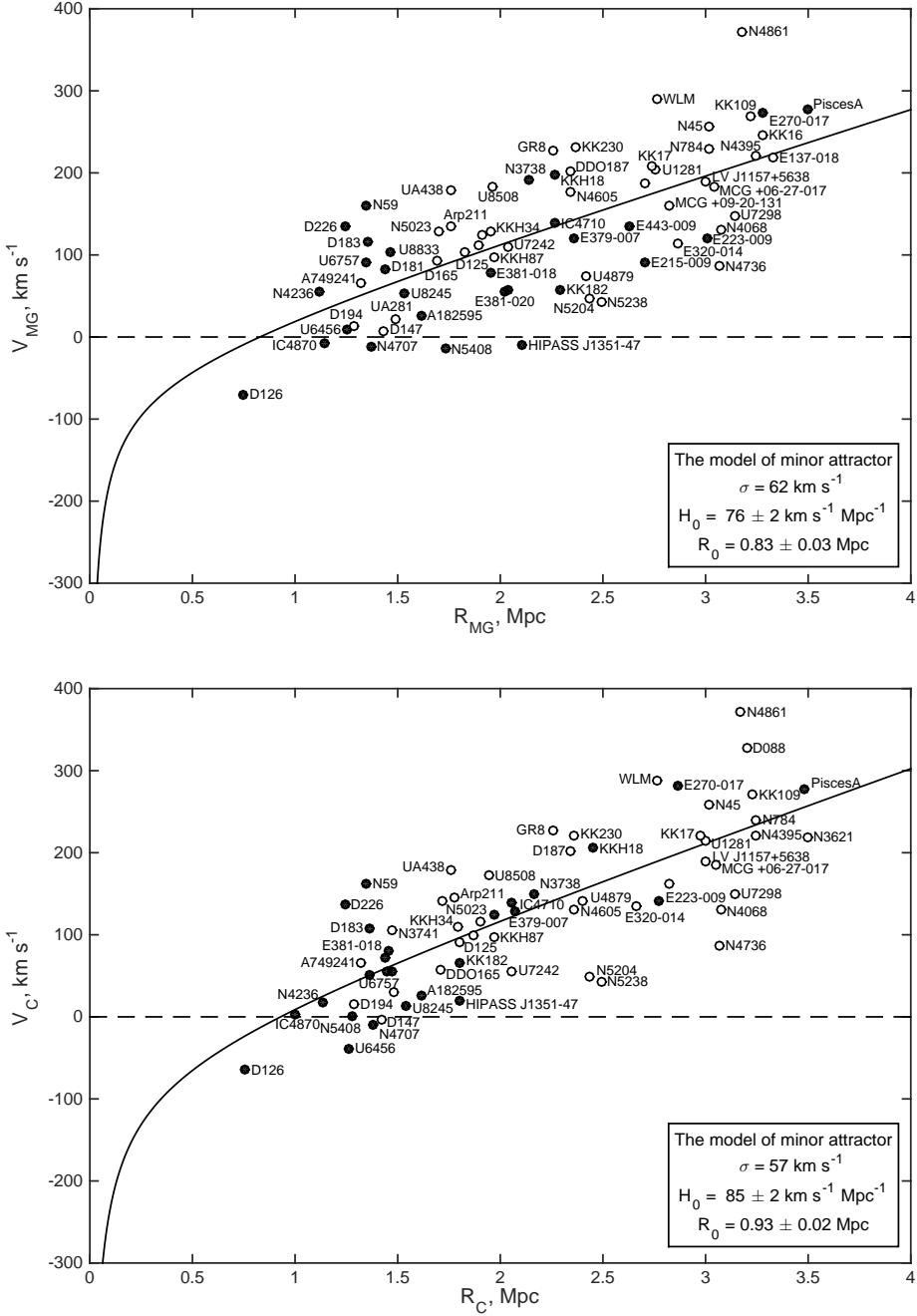
 
\vspace{0.2cm}%
\includegraphics[width=12cm]{synthetic_regression_lv_main_minor.eps}\\%
\vspace{0.2cm}\\%
\sidecaption%
\includegraphics[width=12cm]{synthetic_regression_lv_binary_minor.eps}
\caption{Hubble diagram for the synthetic group of the Local Volume, assuming
the minor attractor model. {\itshape{}Upper panel:} The distances and velocities
of satellites are calculated relative to the main galaxy in a group.
{\itshape{}Lower panel:} The distances and velocities of satellites are
calculated relative to the barycentre of a pair of the most massive galaxies in
each group.\vspace{0.8cm}}
\label{fig5}
\end{figure*}

Seeking to use as much information as possible about companion motions around
the nearby massive galaxies outside their virial zones, we combined the data on
companions of various galaxies into the single synthetic group. To be included
into the consolidated group, a galaxy should satisfy the following four
conditions: (1) a companion has accurate estimates of distance and radial
velocities; (2) the companion distance from the main galaxy, $R_{MG}$, is less
than 3.5 Mpc; 3) the companion belongs to field galaxies, having $TI<0$; and 4)
the companion has a proper aspect, when its position angle $\lambda$ between the
vector of companion radial velocity and the line joining it with the main galaxy
(see Figure~3) lays within $\lambda<45^{\circ}$ or $\lambda>135^{\circ}$.

These conditions are satisfied for 66 galaxies of the Local Volume; the
corresponding data are presented in Table~7. Its columns contain: (1) name of
the main galaxy acting as the centre of its suite; (2) name of a companion
galaxy; (3) companion galaxy distance from the group barycentre; (4, 5)
companion galaxy velocity relative to the group barycentre in the case of minor
or major attractor; (6, 7) tidal index of the galaxy and the name of its main
disturber; and (8) position angle of the companion as indicated in Figure~3.

The Hubble diagram for the synthetic group of the Local Volume for the minor
attractor model with distances and velocities calculated relative to the main
galaxy is shown in the upper panel of Figure~5. The cosmic flow around the
synthetic group is characterized by the Hubble parameter $H_0=(76\pm2)$
km\,s$^{-1}$/Mpc, velocity dispersion $\sigma_v=62$ km\,s$^{-1}$, and radius of
the zero velocity surface $R_0=0.83\pm0.03$ Mpc. As one can see, the radius
$R_0$ turned out to be quite small, corresponding to the effective mass of the
synthetic group of $\sim1.1\times10^{12}M_{\odot}$. To estimate how various
factors influence $R_0$, we constructed another series of Hubble diagrams.
An alternative Hubble diagram with distances and velocities calculated relative
to the group barycentre rather than from the main galaxy itself is presented in
the lower panel of Figure~5. The barycentre is supposed to lie between the two
most massive galaxies of each group given in Table~6. In this case the local
Hubble parameter is $H_0=(85\pm2)$ km\,s$^{-1}$ Mpc, peculiar velocity
dispersion  $\sigma_v=57$ km\,s$^{-1}$, and radius $R_0$ reaches
$R_0=0.93\pm0.02$ Mpc.

As follows from the data on Table~6, the nearby galaxy groups differ
substantially in their stellar and virial masses, $M^*$ and $M_{orb}$, which can
lead to a systematic bias in the averaged $R_0$ estimate. To verify this effect,
we normalized distances of companions around each group to its individual
radius $R_0$, assuming $R_0\propto M_*^{1/3}$ or $R_0\propto M_{orb}^{1/3}$.
After that we did not find any decrease in peculiar velocity dispersion in the
synthetic Hubble diagram.

The resulting values of $H_0, \sigma_v$ and $R_0$ parameters for all discussed
cases are presented in Table~8; i.e. distances and velocities calculated
relative to the  main galaxy or group barycentre and the minor or major
attractor model. These data allow us to conclude, first, that changing a model
from a minor attractor to major attractor increases the $R_0$ estimate and
causes a significant increment in  dispersion, and, second, that accounting for
the second most massive galaxy in a group leads to a notable growth of the $R_0$
estimate.
  
\section{Discussion}

\begin{figure*} 
\vspace{0.2cm}\sidecaption%
\includegraphics[width=12cm]{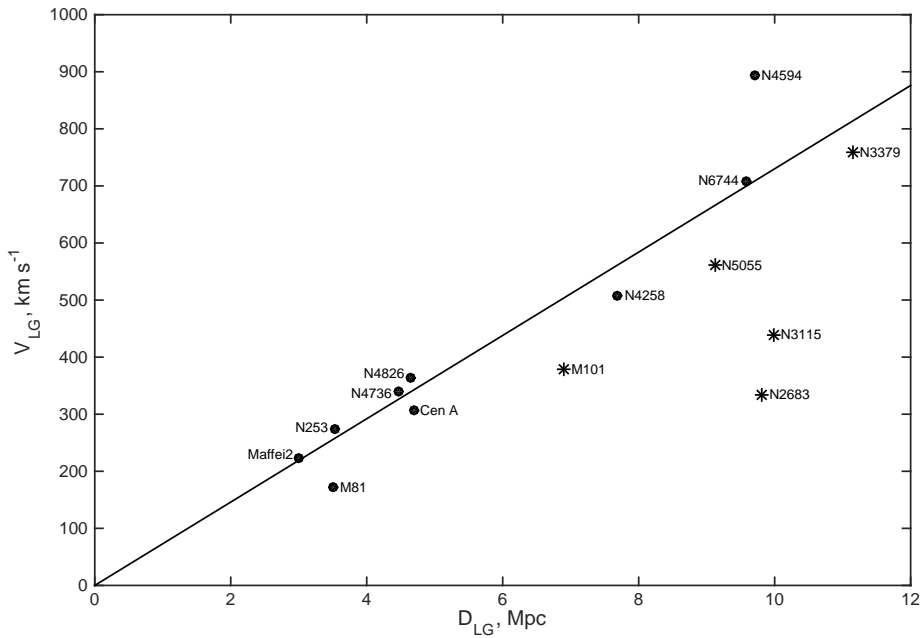}
\caption{Distribution of barycentres of 15 nearby groups by their radial
velocities and distances from the Local Group centre. The solid line corresponds
to the regular Hubble flow with Hubble parameter $H_0
=73$~km\,s$^{-1}$Mpc$^{-1}$.\vspace{0.8cm}}
\label{fig6}
\end{figure*}

As our estimates suggest, galaxies in the infall zone between the virial radius
and the $R_0$ are relatively small in number, $\sim15$\%. This circumstance,
inherent for the Local Group and for other nearby groups, opts for
estimating $R_0$ value within minor attractor model. The low value of peculiar
velocity dispersion resulting in this case is also an oblique argument for such
a choice.

The second most massive galaxy plays an essential role in the kinematics of
several nearby groups, often forming a dynamical subsystem. So, deciding on a
barycentre of the two most bright galaxies as the reference point for distances
and velocities of companions seems to be more preferable than the main galaxy
itself. Hence, we adopt the value of $0.93\pm0.02$ Mpc as the optimal estimate
for $R_0$ radius of the cumulative group (see the lower panel of Figure~5). The
corresponding mass is $\log(M_T/M_{\odot})=12.20$ with a formal error of
$\sim0.04\,$dex. Averaging orbital mass estimates from Table~6 and considering
the representation of each group in the Hubble diagram, we obtain the mean
logarithmic mass $\log(M_{orb}/M_{\odot}) =12.42\pm0.07$. So, the mass of the
synthetic group derived from outer motions of surrounding galaxies turned out to
be $\sim$60\% of the expected mass from inner orbital motions of satellites. A
probable source of this discrepancy was discussed by Chernin et al. (2013) and
Karachentsev \& Kudrya (2014). 

As noted by Chernin et al. (2013), the estimate of the total
mass of a group includes two components,
$M_T = M_m + M_{DE}$, where $M_m$ is the mass of dark and baryonic matter
and $M_{DE}$ is the mass, negative in magnitude,  which is determined by the dark
energy with the density of  $\rho_{DE}$,

\begin{equation}
M_{DE}= (8\pi/3)\times\rho_{DE}\times R^3.
\end{equation}
On the scale of virial radius, the contribution of this component in the group
mass does not exceed 1\%, but in the sphere of $R_0$  radius, the role of this
kind of a mass defect becomes significant. In the standard $\Lambda$CDM  model
with $\Omega_m =0.24$ and $H_0 =73$ km\,s$^{-1}$ Mpc$^{-1}$ the contribution of
dark energy is

\begin{equation}
(M_{DE}/M_{\odot})= -0.85\times 10^{12} \times (R_0/ {\rm Mpc})^3,
\end{equation}
i.e. about 30\% of the Local Group mass determined by orbital motions. This
correction essentially reduces the observed discrepancy between the mass
estimates for the Local Group, as well as for other nearby groups, derived via
internal (virial) and external galaxy motions.

Another possible explanation might be caused by the existence of unrelaxed
(tidal) thin planar structures of satellites seen around the Milky Way and M~31
(Kroupa 2014), which are at variance with the assumption of spherical symmetry
case.

The peculiar velocity dispersion in the vicinity of the synthetic group, 57
km\,s$^{-1}$, is twice as large as in the outskirts of the Local Group. This
difference might originate from bulk motions of galaxies, which become
perceptible on the scale of $\sim5-10$ Mpc. A giant galaxy is not necessarily
the main disturber for neighbouring field galaxies. Indeed, this is the case for
only a portion of companion objects presented in Table~7. Another portion, which
are comprised of mostly distant field galaxies (shown by open circles), are
gravitationally influenced by a massive galaxy from another neighbouring group. 

Figure~6 reproduces the distribution of barycentres of 15 nearby groups listed
in Table~5 by distances and radial velocities relative to the Local Group
centre. The straight line corresponds to the regular Hubble flow with $H_0 =73$
km\,s$^{-1}$ Mpc$^{-1}$. Barycentres of the groups with $D_{LG}<6$ Mpc situated
in the supergalactic plane (i.e. in the Local Sheet; Tully et al. 2016)
demonstrate a small scatter of radial velocities. More distant groups, around
M~101, NGC~5055, NGC~2683, NGC~3115, and NGC~3379 at supergalactic latitudes
$|SGB|>10^{\circ}$ (denoted with asterisks), exhibit negative peculiar
velocities about $-200$~km\,s$^{-1}$. These velocities are caused by the
observed expansion of the Local Void with an amplitude of $\sim
260$~km\,s$^{-1}$(Tully et al. 2016). Also, the group around Sombrero (NGC~4594)
is located just near the zero-velocity surface of the Virgo cluster. Its
positive peculiar velocity reflects the group fall towards the cluster.
Apparently, some portion of these bulk motions manifest themselves as extra
peculiar velocities of the Local Volume galaxies in the panels of Figure~5.
Ignoring these non-virial coherent motions may lead to the overestimation of
galaxy masses based on the Numerical Action Method (Peebles, 2017).

Our conclusion that the peripheral regions of the Local Group and other
neighbouring groups do not contain a large amount of dark matter seems to be the
most important result of this work. The bulk of mass is concentrated within the
virial radius of these groups. The same inference was made for the nearest Virgo
cluster (Karachentsev et al. 2014) from the observed infall of galaxies towards
the cluster centre. Yet further evidence is provided by Kourkchi \& Tully
(2017), who have considered infall zones and collapsed cores of halos in the
Local Universe.

A review of available observational data on distances and radial velocities of
the Local Volume galaxies shows that the population of outskirts of the nearby
groups has not yet been covered with highly accurate distance measurements.
There are groups, for example around the giant Sombrero galaxy, totally lack
reliable distance estimates, even for close probable satellites. The
systematical measurements of TRGB distances with the Hubble Space Telescope
within the Local Volume have the potential to provide meaningful data on the
distribution of the dark matter on the scales of $\sim1$ Mpc.

\begin{acknowledgements}
The authors are grateful to Elena Kaisina for updating the Local Volume Galaxies
Database (https://www.sao.ru/lv/lvgdb/), Dmitry Makarov for the renewed data on
galaxy tidal indices, and Brent Tully for reviewing the manuscript and for his
helpful comments. This work is supported by the Russian Science Foundation grant
No. 14--12--00965.
\end{acknowledgements}
    
{}

\renewcommand{\baselinestretch}{0.8}

\onecolumn
\input{table1.tex}

\onecolumn
\input{table2.tex}

\onecolumn
\input{table3.tex}
\input{table4.tex}
\input{table5.tex}

\onecolumn
\input{table6.tex}

\onecolumn
\input{table7.tex}

\onecolumn
\input{table8.tex}

\end{document}

%% file: table1.tex
\begin{table}
\caption{Milky Way companions with $TI > -0.5$.}
\begin{tabular}{lcrccrr} \\
\hline
 Name        &  RA (2000.0) Dec & $TI$ &   MD &   $D_{MW}$ &   $V_h$ &  $V_{MW}$\\
        & hh\,mm\,ss~dd\,mm\,ss &      &      &   Mpc      & km/s    & km/s    \\
\hline
(1)          &     (2)          & (3)  &  (4) &     (5)    &    (6)  &       (7)\\
\hline
SMC           &005238.0  $-$724801  &3.32 &  LMC &  0.06 &  158  &  17  \\
Sculptor      &010009.4  $-$334233 & 2.79 &  MWay & 0.09 &  105  &  72\\
Phoenix       &015106.3  $-$442641 & 0.73  & MWay & 0.44 &    $-$13 &   $-$103  \\ 
Triangulum II &021317.4  $+$361042 & 3.97 &  MWay & 0.03 &   $-$382  &  $-$257\\
Segue 2       &021916.0  $+$201031 & 3.83 &  MWay & 0.03 &    $-$39 &   44\\
Fornax        &023954.7  $-$343133  &2.19 &  MWay & 0.14 &   29  &   $-$59\\
Horologium 1  &025531.7  $-$540708 & 2.94  & MWay & 0.08 &  113  &   $-$26\\
Reticulum 2   &033542.1  $-$540257 & 4.15  & MWay & 0.03 &   64  &   $-$91\\
Eridanus 2    &034421.1  $-$433159 & 1.00 &  MWay & 0.36  &  76  &   $-$66\\
LMC           &052334.6  $-$694522 & 3.56  & MWay & 0.05 &  278  &  84\\
Carina        &064136.7  $-$505758 & 2.63  & MWay & 0.10 &  224  &   $-$52\\
UMa II        &085130.0  $+$630748 & 3.92 &  MWay & 0.03 &   $-$116  &   $-$33\\
Leo T         &093453.4  $+$170305 & 0.77 &  MWay & 0.42 &   39  &   $-$57\\
Leo A         &095926.4  $+$304447 & 0.03  & MWay & 0.74  &  24  &   $-$17\\
Segue 1       &100703.2  $+$160425 & 4.32 &  MWay & 0.02 &  206 &  111\\
Leo I         &100826.9  $+$121829 & 1.37  & MWay & 0.26 &  283 &  175\\
Sex dSph      &101303.0  $-$013652 & 2.74 &  MWay & 0.09 &  227 &   75\\
UMa I         &103452.8  $+$515512 & 2.59  & MWay & 0.10  &   $-$55  &    $-$7\\
Willman 1     &104921.0  $+$510300 & 3.76 &  MWay & 0.04 &    $-$12  &  36\\
Leo II        &111329.2  $+$220917  &1.45 &  MWay & 0.25 &   86  &  32\\
Leo V         &113109.6  $+$021312 & 1.89  & MWay & 0.18  & 173  &  59\\
Leo IV        &113257.0  $-$003200 & 2.05 &  MWay & 0.16  & 132  &  10\\
Crater        &113615.8  $-$105240 & 1.96 &  MWay & 0.17 &  148  &    $-$2\\
Crater 2      &114914.4  $-$182447 & 2.50 &  MWay & 0.12  &  88  &   $-$74\\
Hydra II      &122142.1  $-$315907 & 2.29 &  MWay & 0.13  & 303  & 129\\
Coma I        &122659.0  $+$235415 & 3.74 &  MWay & 0.04  &  98  &  82\\
CVn II        &125710.0  $+$341915 & 2.04 &  MWay & 0.16  &  $-$129  &   $-$96\\
CVn I         &132803.5  $+$333321  &1.60 &  MWay & 0.22  &  31  &  78\\
Bootes III    &135707.4  $+$264630 & 3.67 &  MWay & 0.05  & 198  & 240\\
Bootes II     &135800.0  $+$125100 & 3.86 &  MWay & 0.04  &  $-$117  &  $-$117 \\
Bootes I      &140000.0  $+$143000 & 3.25 &  MWay & 0.07  &  99  & 106\\
UMin          &150911.3  $+$671252 & 3.27 &  MWay & 0.06 &   $-$255  &   $-$93\\
Hercules      &163102.0  $+$124730 & 2.19 &  MWay & 0.15 &   45  & 145\\
Draco         &172001.4  $+$575434 & 2.94 &  MWay & 0.08 &   $-$296  &  $-$101 \\
Sag dSph      &185503.1  $-$302842 & 5.36 &  MWay & 0.02 &  140  & 169\\
Sag dIr       &192959.0  $-$174041 &  $-$0.44 & MWay & 1.08  &   $-$79   &  7\\
NGC 6822      &194457.7  $-$144811 & 0.52  & MWay & 0.52 &    $-$57  &  43\\
DDO 210       &204651.8  $-$125053 &  $-$0.31 & MWay & 0.98 &   $-$140  &   $-$28\\
Pegasus III   &222424.2  $+$052436 & 1.73 &  MWay & 0.21 &   $-$223 &    $-$63\\
Aquarius 2    &223355.5  $-$091939 & 2.56 &  MWay & 0.11&     $-$71  &  41\\
Tucana        &224149.0  $-$642512 &  $-$0.24 &  MWay & 0.92&   194  &  99\\
Tucana 2      &225155.1  $-$583408 & 3.46  & MWay & 0.06 &   $-$129 &   $-$205 \\
Grus 1        &225642.4  $-$500948 & 2.45 &  MWay & 0.12 &   $-$140 &   $-$186\\
Pisces II     &225831.0  $+$055709 & 1.88 &  MWay & 0.18 &   $-$226  &   $-$75\\
Tucana III    &235636.0  $-$593600 & 3.90 &  MWay & 0.03 &   $-$102 &   $-$195\\
\hline
\end{tabular}
\end{table}

%% file: table2.tex
\begin{table}
\caption{M31 companions.}
\begin{tabular}{lcccccrccr}
\hline
Name        & RA (2000.0) Dec  &  $TI$ & MD &$D_{MW}$& $V_h$   &$D_{M31}$& $R_p$& $\Delta V$\rule{0pt}{3mm}\\
       & hh\,mm\,ss~dd\,mm\,ss &       &    &   Mpc  &  km/s   &   Mpc   &  Mpc & km/s  \\
\hline
(1)         &       (2)        &  (3)  &(4) &   (5)  & (6)  &   (7)   &  (8) &    (9)    \\
\hline
WLM         &000158.1 $-$152740&$-$0.01&M 31 &0.98    &$-$122&0.88     &0.75  &    14     \\
And XVIII   &000214.5 $+$450520&   0.72&M 31 &1.31    &$-$332&0.51     &0.11  & $-$14     \\
And XIX     &001932.1 $+$350237&   2.21&M 31 &0.93    &$-$111&0.16     &0.10  &   187     \\
IC 10       &002024.5 $+$591730&   1.58&M 31 &0.79    &$-$346&0.26     &0.25  & $-$32     \\
And XXVI    &002345.6 $+$475458&   2.58&M 31 &0.76    &$-$261&0.12     &0.10  &    50     \\
Cetus       &002611.0 $-$110240&   0.27&M 31 &0.79    &$-$87 &0.71     &0.69  &    55     \\
And XXV     &003008.9 $+$465107&   3.01&M 31 &0.81    &$-$108&0.09     &0.08  &   200     \\
NGC 147     &003311.6 $+$483028&   2.55&M 31 &0.76    &$-$193&0.12     &0.10  &   115     \\
And III     &003533.8 $+$362952&   2.80&M 31 &0.75    &$-$346&0.10     &0.07  & $-$54     \\
Cas III     &003559.4 $+$513335&   2.28&M 31 &0.78    &$-$372&0.15     &0.14  & $-$64     \\
And XXX     &003634.9 $+$493848&   2.28&NGC 185 &0.68    &$-$141&0.18     &0.11  &   166     \\
And XVII    &003707.0 $+$441920&   2.95&M 31 &0.74    &$-$251&0.09     &0.04  &    51     \\
And XXVII   &003727.1 $+$452313&   3.47&M 31 &0.83    &$-$535&0.06     &0.06  &$-$232     \\
NGC 185     &003858.0 $+$482010&   2.03&M 31 &0.66    &$-$203&0.18     &0.10  &   102     \\
NGC 205     &004022.5 $+$414111&   4.68&M 31 &0.80    &$-$221&0.02     &0.01  &    77     \\
M 32        &004242.1 $+$405159&   4.38&M 31 &0.79    &$-$202&0.03     &0.01  &    93     \\
M 31        &004244.5 $+$411609&   2.79&NGC 205 &0.78    &$-$296&0.00     &0.00  &     0     \\
And I       &004540.0 $+$380214&   2.77&M 31 &0.73    &$-$376&0.10     &0.04  & $-$86     \\
And XI      &004620.0 $+$334805&   2.43&M 31 &0.73    &$-$419&0.14     &0.10  &$-$137     \\
And XII     &004727.0 $+$342229&   2.82&M 31 &0.83    &$-$556&0.10     &0.09  &$-$274     \\
Bol 520     &005042.4 $+$325459&   1.79&M 31 &0.63    &$-$312&0.22     &0.12  & $-$34     \\
And XIV     &005135.0 $+$294149&   2.04&M 31 &0.73    &$-$481&0.18     &0.16  &$-$211     \\
And XIII    &005151.0 $+$330016&   2.55&M 31 &0.84    &$-$195&0.12     &0.12  &    82     \\
And IX      &005252.8 $+$431200&   3.65&M 31 &0.79    &$-$216&0.05     &0.04  &    77     \\
PAndAS-48   &005928.2 $+$312910&   2.31&M 31 &0.82    &$-$250&0.15     &0.14  &    19     \\
And XVI     &005929.8 $+$322236&   1.32&M 31 &0.52    &$-$385&0.32     &0.13  &$-$114     \\
LGS 3       &000355.0 $+$215306&   1.37&M 31 &0.65    &$-$286&0.30     &0.27  & $-$44     \\
IC 1613     &000447.8 $+$020800&   0.64&M 31 &0.76    &$-$232&0.54     &0.53  & $-$59     \\
And X       &000633.7 $+$444816&   1.90&M 31 &0.63    &$-$164&0.20     &0.08  &   124     \\
And V       &001007.1 $+$473741&   2.64&M 31 &0.81    &$-$403&0.11     &0.11  &$-$113     \\
And XV      &001418.7 $+$380703&   2.64&M 31 &0.76    &$-$323&0.11     &0.09  & $-$49     \\
And II      &001629.8 $+$332509&   1.82&M 31 &0.65    &$-$194&0.21     &0.14  &    69     \\
And XXIV    &001830.0 $+$462158&   1.65&M 31 &0.60    &$-$128&0.24     &0.11  &   156     \\
And XXII    &002740.0 $+$280525&   1.75&M 31 &0.79    &$-$127&0.23     &0.22  &   116     \\
And XXIII   &002921.8 $+$384308&   2.24&M 31 &0.73    &$-$243&0.15     &0.13  &    23     \\
M 33        &003350.8 $+$303937&   1.63&M 31 &0.93    &$-$182&0.25     &0.20  &    63     \\
Perseus I   &030123.6 $+$405918&   1.14&M 31 &0.79    &$-$326&0.36     &0.35  &$-$116     \\
And XXVIII  &223241.2 $+$311258&   1.04&M 31 &0.65    &$-$331&0.39     &0.38  &  $-$3     \\
Lac I       &225816.3 $+$411728&   1.50&M 31 &0.76    &$-$198&0.27     &0.26  &   137     \\
Cas dSph    &232631.8 $+$504032&   1.73&M 31 &0.82    &$-$307&0.23     &0.22  &    24     \\
Pegasus     &232834.1 $+$144448&   0.73&M 31 &0.97    &$-$184&0.50     &0.42  &    89     \\
Peg dSph    &235146.4 $+$243510&   1.48&M 31 &0.82    &$-$345&0.28     &0.27  & $-$55     \\
And XXI     &235447.7 $+$422815&   2.42&M 31 &0.86    &$-$361&0.14     &0.12  & $-$43     \\
And XXIX    &235855.6 $+$304520&   1.87&M 31 &0.73    &$-$194&0.21     &0.19  &   106     \\
\hline
PAndAS $-$04&000442.9 $+$472142&   2.5 &M 31 &0.78    &$-$397& ...     &0.12  & $-$79     \\
PAndAS $-$05&000524.1 $+$435535&   2.8 &M 31 &0.78    &$-$183& ...     &0.10  &   132     \\
PAndAS $-$50&010150.6 $+$481819&   2.7 &M 31 &0.78    &$-$323& ...     &0.11  & $-$29     \\
PAndAS $-$56&012303.5 $+$415511&   2.7 &M 31 &0.78    &$-$239& ...     &0.10  &    36     \\
PAndAS $-$57&012747.5 $+$404047&   2.6 &M 31 &0.78    &$-$186& ...     &0.12  &    84     \\
PAndAS $-$58&012902.1 $+$404708&   2.5 &M 31 &0.78    &$-$167& ...     &0.12  &   103     \\
PAndAS $-$01&235712.0 $+$433308&   2.6 &M 31 &0.78    &$-$333& ...     &0.12  & $-$15     \\
PAndAS $-$02&235755.6 $+$414649&   2.6 &M 31 &0.78    &$-$266& ...     &0.11  &    50     \\
\hline
\end{tabular}
\end{table}

%% file: table3.tex
\begin{table}
\caption{Total mass estimates for the Milky Way and M31 (in $10^{12} M_{\sun}$).}
\begin{tabular}{lllcl}
\hline
      M(MW)     &      M(M31)     &  M(MW+M31)  & M(M31)/M(MW) & Reference\\
\hline
$0.75\pm0.25$   & ...             & ...         & ...          & (1) \\
...             & ...             & $3.2\pm0.6$ & ...          & (2) \\
...             & $1.35\pm0.20$   & ...         & ...          & (3) \\
$0.80\pm0.50$   & $1.70\pm0.30$   & $2.5\pm0.6$ & $2.3$        & (4) \\
$0.7\hspace{0.5em}\pm0.4\hspace{0.5em}$ & ...   & ...   & ...  & (5) \\
$1.6\hspace{0.5em}\pm0.4\hspace{0.5em}$ & $1.8\hspace{0.5em}\pm0.5\hspace{0.5em}$ & $3.4\pm0.6$ & $1.1$        & (6) \\
$1.35\pm0.47$   & $1.76\pm0.33$   & $3.1\pm0.6$ & $1.3$        & (7) \\
$1.2\hspace{0.5em}\pm0.4\hspace{0.5em}$ & $0.9\hspace{0.5em}\pm0.3\hspace{0.5em}$ & $2.1\pm0.5$ & $0.75$       & (8) \\
$0.70\pm0.51$   & $1.39\pm0.26$   & $2.1\pm0.6$ & $2.0$        & (9) \\
...             & ...             & $2.6\pm0.4$ & ...          & (10)\\
$1.30\pm0.30$   & ...             & ...         & ...          & (11)\\
$1.02\pm0.76$   & ...             & ...         & ...          & (12)\\
$1.55\pm0.35$   & ...             & ...         & ...          & (13)\\
$2.84$          & $1.65$          & $4.5$       & $0.58$       & (14)\\
$1.18\pm0.18$   & $1.69\pm0.25$   & $2.9\pm0.3$ & $1.4$        & (15)\\
\hline
1.2             & 1.7             & 2.9         & 1.3          & median\\
\hline
\end{tabular}
\end{table}

References. (1) Deason et al. 2012; (2) van der Marel et al. 2012; (3)
Veljanoski et al. 2013; (4) Diaz et al. 2014; (5) Bhattacharjee et al. 2014; (6)
Shull 2014; (7) Karachentsev \& Kudrya 2014; (8) Penarrubia et al. 2014; (9)
Sofue 2015; (10) Penarrubia et al. 2016; (11) McMillan 2017; (12) Patel et al.
2017; (13) Fragione \& Loeb 2017; (14) Peebles 2017; (15) present paper.

%% file: table4.tex
\begin{table}
\caption{Isolated galaxies around the Local Group.}
\begin{tabular}{lcrcrrclr} \hline
Name          &$D_{MW}$& $V_h$&$R_c$&$V_c^{mi}$&$V_c^{ma}$&  TI   &   MD  &$\lambda$\\
              &   Mpc  & km/s & Mpc &   km/s   &   km/s   &       &       & deg     \\
\hline
(1)           &  (2)   & (3)  &(4)  &  (5)     &   (6)    &  (7)  &   (8) &    (9)  \\
\hline
WLM           & 0.98   &$-$122&0.83 & $-$10    &  $-$7    &$-$0.01&M 31    &     27  \\
NGC 404       & 2.98   & $-$50&2.53 &   205    &    205   &$-$0.76&Maffei2 &      1  \\
KKs3          & 2.00   &   316&2.24 &   103    &    109   &$-$1.25&MWay    &     11  \\
KKH 37        & 3.44   &    11&3.17 &   217    &    221   &$-$0.04&M 81    &      6  \\
UGC 4879      & 1.37   & $-$25&1.34 &    25    &     33   &$-$0.63&M 31    &     19  \\
Leo A         & 0.74   &    24&0.93 & $-$47    &  $-$53   &   0.03&MWay    &     29  \\
Sex B         & 1.43   &   300&1.71 &    94    &    101   &$-$0.82&MWay    &     13  \\
NGC 3109      & 1.34   &   403&1.73 &    94    &     96   &$-$0.33&Antlia  &      9  \\
Sex A         & 1.45   &   324&1.78 &    78    &     82   &$-$0.83&MWay    &     11  \\
Leo P         & 1.73   &   262&1.95 &   120    &    128   &$-$1.07&MWay    &     12  \\
NGC 3741      & 3.22   &   229&3.27 &   255    &    262   &$-$0.69&M 81    &      8  \\
DDO 99        & 2.65   &   251&2.76 &   247    &    255   &$-$0.62&NGC 4214 &      9  \\
IC 3104       & 2.36   &   429&2.73 &   159    &    162   &$-$1.12&NGC 4945 &      6  \\
DDO 125       & 2.61   &   206&2.69 &   242    &    251   &$-$0.94&M 81    &     10  \\
DDO 147       & 3.01   &   331&3.14 &   342    &    350   &$-$0.60&NGC 4214 &      8  \\
GR 8          & 2.19   &   217&2.48 &   128    &    132   &$-$1.37&MWay    &      9  \\
UGC 8508      & 2.67   &    56&2.66 &   176    &    184   &$-$0.80&M 81    &     10  \\
DDO 181       & 3.10   &   214&3.19 &   278    &    285   &$-$0.87&NGC 4736 &      8  \\
DDO 183       & 3.31   &   188&3.41 &   247    &    253   &$-$0.79&NGC 4736 &      8  \\
KKH 86        & 2.61   &   287&2.93 &   198    &    202   &$-$1.38&NGC 5128 &      7  \\
UGC 8833      & 3.25   &   221&3.37 &   273    &    280   &$-$0.89&NGC 4736 &      8  \\
KK 230        & 2.21   &    63&2.34 &   120    &    127   &$-$1.34&M 81     &     11  \\
DDO 187       & 2.30   &   160&2.51 &   171    &    178   &$-$1.44&MWay     &     10  \\
DDO 190       & 2.83   &   150&2.88 &   258    &    267   &$-$1.18&M 81     &      9  \\
ESO 274--01   & 2.79   &   524&3.20 &   327    &    329   &$-$0.51&NGC 5128 &      4  \\
KKR 25        & 1.91   & $-$79&1.84 &   126    &    137   &$-$0.98&M 31     &     14  \\
IC 4662       & 2.55   &   302&2.87 &   131    &    135   &$-$1.24&NGC 5128 &      7  \\
NGC 6789      & 3.55   &$-$140&3.28 &   153    &    156   &$-$1.32&M 81     &      6  \\
Sag dIr       & 1.08   & $-$79&1.19 &    20    &     29   &$-$0.44&MWay     &     22  \\
DDO 210       & 0.98   &$-$140&0.98 &    12    &     22   &$-$0.31&MWay     &     27  \\
IC 5152       & 1.96   &   122&2.08 &    70    &     77   &$-$1.20&NGC 253  &     12  \\
KK 258        & 2.24   &    92&2.18 &   151    &    160   &$-$0.91&NGC 253  &     12  \\
Tucana        & 0.92   &   194&1.14 &    61    &     77   &$-$0.24&MWay     &     22  \\
UGCA 438      & 2.22   &    62&2.15 &   101    &    108   &$-$0.48&NGC 55   &     12  \\
KKH 98        & 2.58   &$-$132&2.14 &   171    &    171   &$-$0.93&M 31     &      2  \\
\hline
\end{tabular}
\end{table}

%% file: table5.tex
\begin{table}
\caption{Parameters of the local Hubble flow as a function of the LG barycentre
         position.}
\begin{tabular}{lccccccccc}
\hline
$x = D_C/D_{M31}$    & 0.10 & 0.20 & 0.30 & 0.40 & 0.50 & 0.60 & 0.70 & 0.80 & 0.90\\
$M_{M31}/M_{MW}$      & 0.11 & 0.25 & 0.43 & 0.67 & 1.00 & 1.50 & 2.33 & 4.00 & 9.00\\
$\sigma_v, km s^{-1}$ & 26.7 & 24.9 & 23.6 & 23.1 & 23.4 & 24.2 & 25.7 & 27.7 & 30.0\\
$R_0, Mpc $          & 0.83 & 0.85 & 0.87 & 0.89 & 0.91 & 0.93 & 0.95 & 0.97 & 0.99\\
\hline
\end{tabular}
\end{table}

%% file: table6.tex
\begin{table}
\caption{Giant galaxies in the Local Volume.}
\begin{tabular}{lrrlrrcc} \hline

 Galaxy   &  SGL &   SGB  &$D_{MW}$&   $V_{LG}$ &   $\lg M^*$  &$\lg M_{orb}$ & $N_{sat}$\\
          &  deg &   deg  &   Mpc  &    km/s    &   $M_{\odot}$&$M_{\odot}$   &(V,D)\\
\hline
   (1)    &  (2) &   (3)  &  (4) &     (5)    &      (6)     &      (7)     & (8)\\
\hline
\vspace{-1mm}\\
 M 31     &  336.19 &  12.55  &  0.78  &     $-$29  &   10.79 &  12.49 &   90\\
 M Way   &       ...    &     ...    &  0.01  &     $-$65  &   10.70& &\\
\vspace{-1mm}\\
 M 81     &   41.12 &   0.59  &  3.70  &   104  &   10.95 &  12.69 &   22\\
 M 82     &   40.72 &   1.05  &  3.61  &   328  &   10.59& &\\
\vspace{-1mm}\\ 
 NGC 5128 &  159.75 &    $-$5.25  &  3.68  &   310  &   10.89 &  12.89 &   28\\
 NGC 5236 &  147.93 &   0.99  &  4.90  &   307   &  10.86 &  &\\
\vspace{-1mm}\\
 Maffei 2 &  359.58 &   0.83  &  3.48 &    214  &   10.86 &  12.51&     3\\
 IC 342   &   10.60 &   0.37  &  3.28 &    244  &   10.60& &\\
\vspace{-1mm}\\
 NGC 253  &  271.57 &    $-$5.01 &   3.70 &    276  &   10.98 &  12.18 &    7\\
 NGC 247  &  275.92 &    $-$3.73 &   3.72 &    216  &    9.50& &\\
\vspace{-1mm}\\
 NGC 4826 &   95.61 &   6.13 &   4.41 &    365  &   10.49 &  10.78 &    4\\
 DDO 154  &   90.13 &   6.90 &   4.04 &    354  &    7.59& &\\
\vspace{-1mm}\\
 NGC 4736  &  76.24 &   9.50 &   4.41 &    352  &   10.56 &  12.43 &   16 \\
 NGC 4449  &  72.30 &   6.18 &   4.27 &    249  &    9.68& &\\
\vspace{-1mm}\\
 M 101    &   63.58 &  22.61 &   6.95 &    378   &  10.79 &  12.17&     6\\
 NGC 5474 &   64.30 &  22.93 &   6.98 &    424   &   9.21& &\\
\vspace{-1mm}\\
 NGC 4258 &   68.74 &   5.55 &   7.66 &    506  &   10.92 &  12.50&     7\\
 NGC 4242 &   70.28 &   4.81 &   7.9: &    568   &   9.47& &\\
\vspace{-1mm}\\
 NGC 5055 &   76.20 &  14.25 &   9.04 &    562  &   11.00 &  12.49 &    0\\
 NGC 4460  &  71.58 &   6.48 &   9.59 &    551  &    9.66& &\\
\vspace{-1mm}\\
 NGC 4594 &  126.69 &    $-$6.68 &   9.30 &    894  &   11.30 &  13.45 &    0\\
 NGC 4597 &  121.05 &    $-$5.12 &  10.1: &    912  &    9.48& &\\
\vspace{-1mm}\\
 NGC 6744 &  208.10 &  10.38 &   9.51 &    706  &   10.91 & 11.72 &    4\\
 NGC 6684  & 205.81 &   9.11 &   8.7: &    720  &   10.39&&\\
\vspace{-1mm}\\
 NGC 3115 &  112.40 &   $-$42.86 &   9.68 &    439  &   10.95 & 12.54 &    0\\
 P 4078671&  114.10 &   $-$45.34  &  9.38 &    378  &    7.95&&\\
\vspace{-1mm}\\
 NGC 2683  &  55.87 &   $-$33.42 &   9.82 &    334  &   10.81 & 12.13 &    2\\
 KK 69     &  55.64 &   $-$33.09 &   9.16 &    418   &   7.27&&\\
\vspace{-1mm}\\
 NGC 3379  &  93.64 &   $-$25.85 &  11.32 &    774  &   10.92 & 13.23 &    3\\
 NGC 3368  &  94.29 &   $-$26.41 &  10.42 &    740  &   10.83&&\\
\hline
\end{tabular}
\end{table}

%% file: table7.tex
\begin{table}
\caption{Isolated galaxies around the nearby group centres.} 
\begin{tabular}{llcrrrlr}\hline
Main gal.& Name           &$R_C$&$V_{mi}$&$V_{ma}$&   TI   &  MD    &  $\lambda$\\
         &                & Mpc &  km/s  &  km/s  &        &        &   deg  \\
\hline
 (1)     &  (2)           & (3) &   (4)  &  (5)   &   (6)  &  (7)   &  (8)\\
\hline
M81      &UGC04879        &2.40 &  141   &   147  &$-$0.63 &M31     & 153\\
M81      &UGC06456        &1.26 &$-$38   & $-$86  &$-$0.31 &M81     &  35\\
M81      &NGC3738         &2.16 &  150   &   170  &$-$1.01 &M81     &  33\\
M81      &UGC06757        &1.37 &   51   &    42  &$-$0.41 &M81     &  40\\
M81      &UGC07242        &2.06 &   56   &    48  &$-$0.40 &N4605   &  24\\
M81      &NGC4236         &1.13 &   17   & $-$15  &$-$0.16 &M81     &  44\\
M81      &NGC4605         &2.36 &  131   &   138  &$-$1.07 &M101    &  29\\
M81      &DDO165          &1.71 &   56   &    41  &$-$0.64 &N4236   &  39\\
M81      &UGC08245        &1.54 &   14   & $-$27  &$-$0.58 &M81     &  40\\
NGC5128  &NGC3621         &3.50 &  219   &   204  &$-$1.68 &N4594   &  34\\
NGC5128  &ESO320-014      &2.66 &  134   &   100  &$-$0.68 &N3621   &  38\\
NGC5128  &ESO379-007      &2.07 &  129   &   111  &$-$1.04 &N5236   &  45\\
NGC5128  &ESO381-018      &1.45 &   81   &    74  &$-$0.40 &N5236   &  29\\
NGC5128  &ESO381-020      &1.47 &   55   &    39  &$-$0.33 &N5236   &  29\\
NGC5128  &ESO443-009      &1.97 &  124   &   119  &$-$0.53 &N5236   &  24\\
NGC5128  &KK182           &1.80 &   66   &    62  &$-$0.67 &N5236   &  16\\
NGC5128  &ESO270-017      &2.86 &  282   &   289  &$-$1.35 &N5236   &  15\\
NGC5128  &HIPASS J1348-37 &1.45 &   56   &    54  &$-$0.21 &N5236   &  12\\
NGC5128  &HIPASS J1351-47 &1.80 &   20   & $-$11  &$-$0.87 &N5236   &  29\\
NGC5128  &NGC5408         &1.28 &    0   & $-$25  &$-$0.35 &N5236   &  29\\
NGC5128  &ESO223-009      &2.78 &  142   &   120  &$-$1.42 &N5236   &  33\\
Maffei2  &UGC01281        &3.00 &  215   &   223  &$-$1.20 &N784    &  37\\
Maffei2  &KK17            &2.97 &  220   &   238  &$-$0.96 &N784    &  41\\
Maffei2  &NGC0784         &3.25 &  240   &   254  &$-$1.30 &U1281   &  37\\
Maffei2  &KKH18           &2.45 &  207   &   243  &$-$1.17 &Maffei2 &  42\\
Maffei2  &KKH34           &1.80 &  110   &   106  &$-$0.65 &M81     &  39\\
N253     &WLM             &2.76 &  288   &   298  &$-$0.01 &M31     & 161\\
N253     &NGC0045         &3.02 &  259   &   260  &$-$1.05 &N24     &  10\\
N253     &PiscesA         &3.49 &  278   &   292  &$-$1.68 &N253    &  40\\
N253     &NGC0059         &1.35 &  163   &   174  &$-$0.37 &N253    &  23\\
N253     &DDO226          &1.24 &  136   &   138  &$-$0.27 &N253    &   9\\
N253     &UGCA438         &1.76 &  178   &   221  &$-$0.48 &N55     & 136\\
N4826    &AGC749241       &1.32 &   65   &    59  &$-$0.73 &N4656   &  20\\
N4826    &GR8             &2.26 &  227   &   230  &$-$1.37 &MW      & 165\\
N4826    &DDO187          &2.34 &  202   &   207  &$-$1.44 &MW      & 144\\
N4736    &NGC3741         &1.47 &  105   &    95  &$-$0.69 &M81     & 135\\
N4736    &DDO099          &1.87 &   99   &    86  &$-$0.62 &N4214   & 153\\
N4736    &UGCA281         &1.48 &   30   &    12  &$-$0.92 &N4258   &  24\\
N4736    &DDO126          &0.76 &$-$64   &$-$136  &$-$0.02 &N4736   &  37\\
N4736    &DDO125          &1.80 &   92   &    90  &$-$0.94 &M81     & 169\\
N4736    &Arp211          &1.78 &  146   &   146  &$-$0.86 &N4258   &   9\\
N4736    &DDO147          &1.42 & $-$4   & $-$13  &$-$0.60 &N4214   & 164\\
N4736    &NGC5023         &1.72 &  141   &   141  &$-$0.89 &M101    &  13\\
N4736    &UGC08508        &1.94 &  172   &   183  &$-$0.80 &M81     & 144\\
N4736    &DDO181          &1.44 &   73   &    59  &$-$0.87 &N4736   & 149\\
N4736    &DDO183          &1.36 &  107   &   109  &$-$0.79 &N4736   & 136\\
N4736    &KK230           &2.36 &  220   &   235  &$-$1.34 &M81     & 148\\
N4736    &DDO190          &1.91 &  116   &    85  &$-$1.18 &M81     & 135\\
M101     &LV J1157+5638   &3.00 &  190   &   215  &$-$0.80 &N4258   &  45\\
M101     &NGC4068         &3.07 &  131   &    98  &$-$0.48 &N4736   & 137\\
M101     &MCG +09-20-131  &2.82 &  161   &   167  &$-$0.43 &N4736   & 137\\
M101     &UGC07298        &3.14 &  149   &   136  &$-$0.35 &N4736   & 142\\
M101     &NGC4736         &3.07 &   87   &    12  &$-$0.13 &N4449   & 136\\
M101     &NGC5204         &2.44 &   49   &    40  &$-$0.88 &N4736   & 162\\
M101     &NGC5238         &2.49 &   43   &    37  &$-$0.41 &N4736   & 166\\
M101     &KKH87           &1.97 &   97   &    96  &$-$0.81 &N5194   &  11\\
M101     &DDO194          &1.29 &   15   &  $-$4  &$-$0.10 &N5585   & 150\\
N4258    &KK109           &3.23 &  270   &   275  &$-$0.32 &N4736   & 164\\
N4258    &MCG +06-27-017  &3.05 &  185   &   180  &$-$0.22 &N4395   & 151\\
N4258    &NGC4395         &3.25 &  222   &   225  &$-$0.12 &N4736   & 146\\
N4258    &NGC4707         &1.38 & $-$9   & $-$66  &$-$0.45 &N4258   & 143\\
N4258    &NGC4861         &3.17 &  371   &   464  &$-$0.57 &N5055   &  37\\
N6744    &IC4710          &2.06 &  139   &   137  &$-$0.99 &N6684   & 158\\
N6744    &IC4870          &1.00 &    4   & $-$40  &$-$0.22 &N6744   & 142\\
N2683    &AGC182595       &1.61 &   27   &    12  &$-$0.78 &N2683   & 144\\
N3379    &DDO088          &3.20 &  328   &   329  &$-$0.45 &N3627   & 174\\
\hline
\end{tabular}
\end{table}

%% file: table8.tex
\begin{table}
\caption{Parameters of the Hubble flow around the nearby synthetic group
	   under various assumptions.}
\begin{tabular}{rccc|ccc}\hline
\multicolumn{1}{c}{Case}&
\multicolumn{3}{c}{Minor attractor}&
\multicolumn{3}{c}{Major attractor}\\ \hline

		      &  $H_0$ &  $\sigma_v$ &  $R_0$ &      $H_0$ &  $\sigma_v$ &  $R_0$\\
&km\,s$^{-1}$\,Mpc$^{-1}$&km\,s$^{-1}$& Mpc&km\,s$^{-1}$\,Mpc$^{-1}$&km\,s$^{-1}$&Mpc\\
\hline
    (1)	      &  (2)   &      (3)    &  (4)   &      (5)   &     (6)     &   (7) \\
\hline
 Main galaxy (MG)     &  76   &   62   &   0.83    &   80   &   84   &    0.93   \\ 
 MG-normalized        &  80   &   65   &   0.71    &   82   &   89   &    0.76 \\
 Barycenter (BC)      &  85   &   57   &   0.93    &   85   &   84   &    1.03\\
 BC-normalized        &  88   &   66   &   0.76    &   95   &   85   &    0.93\\
\hline 
\end{tabular}
\end{table}

%% file: 31645_corr_v2.bbl
\begin{thebibliography}{}
\bibitem{} Abazajian K.N., Adelman-McCarthy J.K., Ag\"{u}eros M.A., et al. 2009,
   ApJS, 182, 543
\bibitem{} Barber C., Starkenburg E., Navarro J.F.,et al. 2014, MNRAS, 437, 959
\bibitem{} Bhattacharjee P., Chaudhury S., Kundu S., 2014, ApJ, 785, 63
\bibitem{} Chernin A.D., Bisnovatyi-Kogan G.S., Teerikorpi P., et al. 2013,
   A\&A, 553, 101
\bibitem{} Chernin A.D., Karachentsev I.D., Valtonen M.J., et al. 2004, A\&A,
   415, 19
\bibitem{} Deason A.J., Belokurov V., Evans N.W., et al. 2012, MNRAS, 425, 2840
\bibitem{} de Vaucouleurs G., 1958, AJ, 63, 253
\bibitem{} de Vaucouleurs G., 1964, AJ, 69, 737
\bibitem{} de Vaucouleurs G., 1972, in {\itshape{}IAU Symposium 44, External
   Galaxies and Quasi-Stellar Objects}, ed. D.\,S.\,Evans, D.\,Wills, and
   B.\,J.\,Wills (Dordrecht: Reidel), p.\,353
\bibitem{} Diaz J.D., Koposov S.E., Irwin M., et al. 2014, MNRAS, 443, 1688
\bibitem{} Ekholm, T., Baryshev, Y., Teerikorpi, P. et al., 2001, A \& A, 368,
   L17
\bibitem{} Falco M., Hansen S.H., Wojtak R. et al. 2014, MNRAS, 442, 1887
\bibitem{} Fragione, G., \& Loeb, A. 2017, New A, 55, 32
\bibitem{} Huxor A.P., Mackey A.D., Ferguson A.M.N, et al. 2014, MNRAS, 442,
   2165
\bibitem{} Ibata R. A., Lewis G. F., McConnachie A. W., et al., 2014, ApJ, 780,
   128
\bibitem{} Ibata, R.A., Lewis, G.F., Conn, A.R. et al. 2013, Nature, 493, 62
\bibitem{} Ibata R., Martin N. F., Irwin M., et al., 2007, ApJ, 671, 1591
\bibitem{} Kaisina E.I., Makarov D.I., Karachentsev I.D., 2012, AstBull, 67, 115
\bibitem{} Karachentsev I.D., Tully R.B., Wu Po-Feng, Shaya E.J., Dolphin A.E.,
   2014, ApJ, 782, 4
\bibitem{} Karachentsev I.D., Kudrya Y.N., 2014, AJ, 148, 50
\bibitem{} Karachentsev I.D., Makarov, D.I., \& Kaisina, E.I., 2013, AJ, 145,
   101 (=UNGC)
\bibitem{} Karachentsev I.D., Nasonova (Kashibadze) O.G., 2010, MNRAS, 405, 1075
\bibitem{} Karachentsev I.D., Kashibadze O.G., Makarov D.I., Tully R.B., 2009,
   MNRAS, 393, 1265
\bibitem{} Karachentsev I.D., Kashibadze O.G., 2006, Ap, 49,3
\bibitem{} Karachentsev I.D., Dolphin A.E., Tully R.B., et al, 2006, AJ, 131,
   1361
\bibitem{} Karachentsev I.D., Sharina M.E., Makarov D.I., et al, 2002, A\&A,
   389, 812
\bibitem{} Koposov S.E., Belokurov V., Torrealba G., Wyn E.N., 2015, ApJ, 805,
   130
\bibitem{} Kourkchi E., Tully R.B., 2017, ApJ, 843, 16
\bibitem{} Kroupa P., 2014, ArXiv1409.6302
\bibitem{} Lynden-Bell D., 1981, Observatory, 101, 111
\bibitem{} Martin N. F., McConnachie A. W., Irwin M., et al., 2009, ApJ, 705,
   758
\bibitem{} McConnachie A. W., 2012, AJ, 144, 4
\bibitem{} McMillan P.J., 2017, MNRAS, 465, 76
\bibitem{} Nasonova O.G., de Freitas Pacheco J.A., Karachentsev I.D., 2011,
   A\&A, 532A, 104
\bibitem{} Patel E., Besla G., \& Mandel K., 2017, MNRAS, 468, 3428
\bibitem{} Pawlowski, M.S., Famaey, B., Jerjen, H. et al. 2014, MNRAS, 442,
   2362
\bibitem{} Peebles P.\,J.\,E., 2017, arXiv:1705.10683
\bibitem{} Peebles P.\,J.\,E., 1976, ApJ, 205, 318
\bibitem{} Peirani, S., \& de Freitas Pacheco,J.A. 2008, A\&A, 488, 845
\bibitem{} Penarrubia J, Fattahi A., 2017, MNRAS, 468, 1300
\bibitem{} Penarrubia J, Gomez, F.A., Besla G., et al. 2016, MNRAS, 456, L54
\bibitem{} Penarrubia J, Ma Y.Z., Walker M.G., McConnachie A. W., 2014, MNRAS,
   443, 2204
\bibitem{} Planck Collaboration, Ade P.A.R., Aghanim N., et al., 2014, A \& A,
   571, A16
\bibitem{} Sandage A., 1986, ApJ, 307, 1
\bibitem{} Sandage A., Tammann G.\,A., 1975, ApJ, 196, 313
\bibitem{} Shull J.M., 2014, ApJ, 784, 142
\bibitem{} Sofue Y., 2015, PASJ, 67, 75
\bibitem{} Spergel D.N., 2007, ApJS, 170, 377
\bibitem{} Teerikorpi, P., Chernin, A.D., \& Baryshev, Y.V., 2005, A\&A, 440,
   791
\bibitem{} Tonry J. L., Stubbs C. W., Lykke K. R., et al., 2012, ApJ, 750, 99
\bibitem{} Tully R.B., Courtois H.M., Sorce J.G., 2016, AJ, 152, 50
\bibitem{} Tully R.B., Shaya E.J., 1984, ApJ, 281, 31
\bibitem{} Tully R.B., Fisher J.R., 1977, A\&A, 54, 661
\bibitem{} van der Marel R.P., Fardal M., Besla G. et al. 2012, ApJ, 753, 8
\bibitem{} Veljanoski J., Ferguson A.M.N., Mackey A.D., et al. 2013, ApJ, 768L, 33
\end{thebibliography}
